\documentclass[%
 reprint,
 amsmath,amssymb,
 aps,
]{revtex4-2}
\usepackage{graphicx}
\usepackage{dcolumn}
\usepackage{bm}

\begin{document}

\preprint{APS/123-QED}

\title{Metamaterial-Controlled Parity-Time Symmetry in Non-Hermitian Wireless Power Transfer Systems}

\author{Hanwei Wang}
\affiliation{%
 Department of Electrical and Computer Engineering, University of Illinois at Urbana-Champaign, Urbana, IL, US.
}%

 \author{Joshua Yu}%
\affiliation{%
 Department of Electrical and Computer Engineering, University of Illinois at Urbana-Champaign, Urbana, IL, US.
}

\author{Xiaodong Ye}%
\affiliation{%
 Department of Electrical and Computer Engineering, University of Illinois at Urbana-Champaign, Urbana, IL, US.
}%

\author{Yang Zhao}%
 \email{yzhaoui@illinois.edu}
\affiliation{%
 Department of Electrical and Computer Engineering, University of Illinois at Urbana-Champaign, Urbana, IL, US.
}%

\date{\today}

\begin{abstract}
Inductive wireless power transfer (WPT) systems, modeled as non-Hermitian systems using coupled-mode theory, leveraging parity-time (PT)-symmetric states for efficient power transfer. However, traditional passive relay resonators in these systems can induce additional eigenstates with broken PT symmetry due to spatial constraints. Here, we introduce a theory involving a multibody WPT system with metamaterial-controlled PT symmetry, overcoming the limitations and achieving free-positioning WPT. Using inverse design, we configure the metamaterial to target a resonant mode that balances the effective coupling coefficients between the metamaterial, transmitting (Tx), and receiving (Rx) resonators, ensuring a stable PT-symmetric state in a strong coupling regime, confirmed through numerical calculations and experimental validations. Notably, our experiments show PT-symmetric state formation with varying Tx and Rx sizes and positions, as well as different Rx spatial configurations, highlighting our system's potential for versatile WPT applications.
\end{abstract}

\maketitle


\section{Introduction}
Wireless power transfer (WPT) technologies are broadly divided into two main categories, radiative \cite{krasnok2018coherently, ng2014robust} and non-radiative \cite{kurs2007wireless}. Non-radiative WPT, which primarily uses the magnetic near-field to transmit energy, is preferred for its high-power volume and safety features \cite{sasatani2021room}. The transmitting (Tx) and receiving (Rx) resonators couple through magnetic mutual induction \cite{karalis2008efficient}. These inductive WPT systems can be modeled as non-Hermitian systems using the coupled-mode theory \cite{assawaworrarit2017robust}. Efficient power transfer in these systems is achieved by forming PT-symmetric states \cite{schindler2012symmetric, cao2022fully}, especially in the strong coupling regime when the physical symmetry is maintained \cite{assawaworrarit2020robust}. However, a challenge arises in the weak coupling regime typically when the Tx-Rx separation increases. In such scenarios, spontaneous symmetry breaking occurs, leading to the formation of anti-PT-symmetric resonant states \cite{el2018non}. 

To enhance the overall coupling and extend the strong coupling range, relay resonators have been employed to increase the maximum separation distance between Tx and Rx \cite{zhong2011general, wu2022generalized}. However, a significant challenge with relay resonators is their propensity to involve high-order resonant states, many of which exhibit anti-PT symmetry \cite{zeng2020high}. To avoid these states, it is essential to maintain a symmetric spatial arrangement of the relay resonators and ensure identical geometries for the Tx and Rx resonators \cite{hao2023frequency}. Yet, such stringent requirements are impractical in many applications, particularly in scenarios requiring free-positioning WPT. 

Metamaterials, known for their exceptional ability in manipulating the wavefront and near-field distribution of electromagnetic and acoustic fields and waves \cite{shelby2001experimental, lawrence2014manifestation, sounas2015unidirectional, cummer2016controlling,  yu2011light}, hold significant promise in addressing these challenges in WPT systems. Prior research has successfully harnessed metamaterials to control the PT symmetry in photonics, leading to developments like asymmetric phase modulation \cite{mikheeva2023asymmetric}, isotropic negative refractive index \cite{alaeian2014parity}, nanoscale sensing \cite{park2020symmetry, chen2017exceptional, sakhdari2017pt}, and coherent perfect absorption \cite{kang2013effective}. However, applying these concepts to WPT systems is hindered by difficulties in controlling the metamaterial’s resonant mode. In our previous work, we have successfully demonstrated a metamaterial that can achieve on-demand field-shaping, applicable in magnetic resonance imaging \cite{wang2021demand} and WPT \cite{wang2021wearable, wang2023qi}. Here, we show that controlling the metamaterial’s mode fine-tunes the coupling coefficients relative to the Tx and Rx, enabling a PT-symmetric state without needing identical Tx and Rx sizes or positions. This advancement paves ways for new multibody WPT system designs.

\section{Theory}
The circuit diagram illustrating our metamaterial-controlled WPT system is shown in FIG. 1(a). While we demonstrate the concept using a single layer metamaterial (i.e., a metasurface), our theory is general and applicable to metamaterials composing any number of layers. Within this system, the metamaterial functions as a controllable relay to bridge the Tx and Rx resonators. The effective coupling between the metamaterial and the Tx/Rx resonators is determined by the targeted resonant mode ${{\bf{a}}_{\bf{t}}}$ of the metamaterial, ${\kappa _{1m}} = {{\bf{\kappa }}_{{\bf{1u}}}}^{\bf{\dag }}{{\bf{a}}_{\bf{t}}}$ and ${\kappa _{2m}} = {{\bf{\kappa }}_{{\bf{2u}}}}^{\bf{\dag }}{{\bf{a}}_{\bf{t}}}$, where ${{\bf{\kappa }}_{{\bf{1u}}}}$ (or ${{\bf{\kappa }}_{{\bf{2u}}}}$) are the coupling coefficients between the Tx (or Rx) and the metamaterial resonators. Specifically, ${{\bf{\kappa }}_{{\bf{1u}}}} = {\left[ {\begin{array}{*{20}{c}}
{{\kappa _{1u1}}}& \cdots &{{\kappa _{1uN}}}
\end{array}} \right]^T}$
, with ${\kappa _{1uj}}$ being the coupling coefficient between the Tx resonator and the metamaterial’s j-th unit cell. 
${\kappa _0}$ is the coupling between the Tx and Rx resonators.

\begin{figure}[b]
\includegraphics[width=0.48\textwidth]{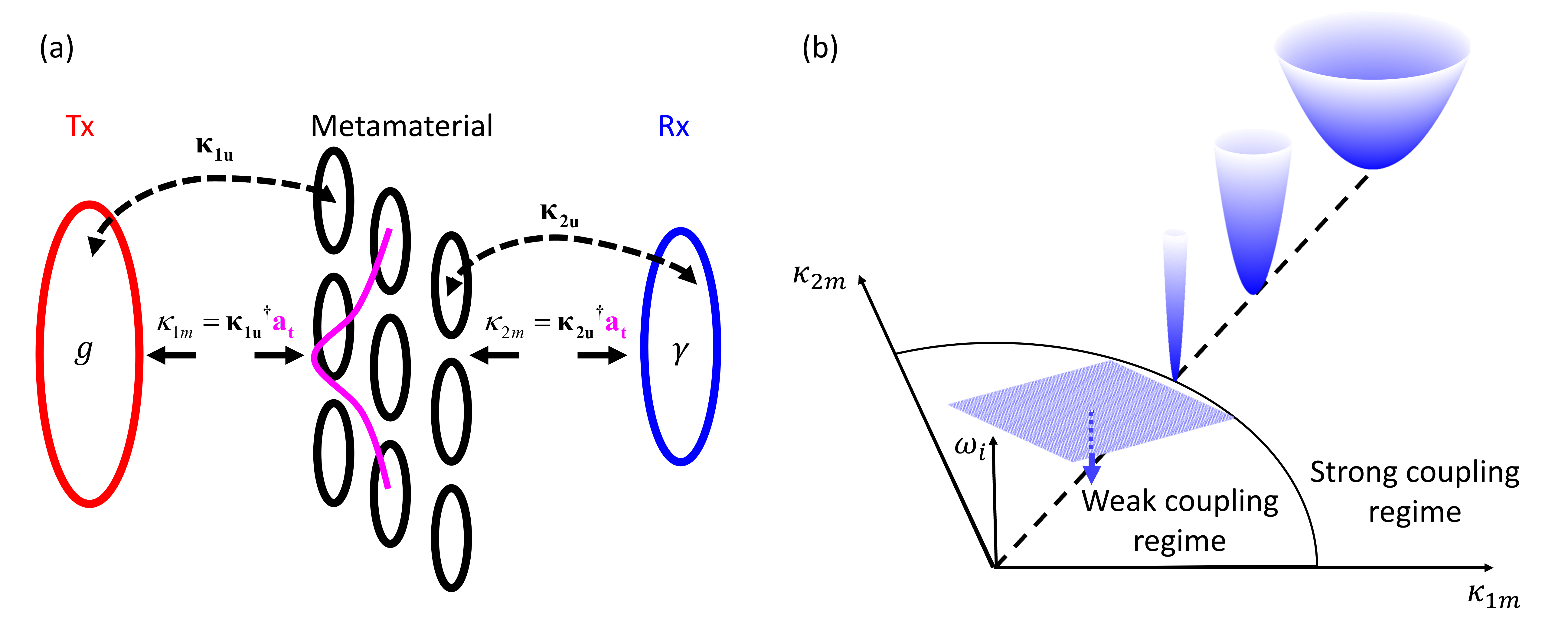}
\caption{\label{fig:epsart} Metamaterial-enhanced WPT system. (a) The Tx and Rx resonators are coupled to the metamaterial. The coupling coefficients are given by ${\kappa _{1m}} = {{\bf{\kappa }}_{{\bf{1u}}}}^{\bf{\dag }}{{\bf{a}}_{\bf{t}}}$ and ${\kappa _{2m}} = {{\bf{\kappa }}_{{\bf{2u}}}}^{\bf{\dag }}{{\bf{a}}_{\bf{t}}}$, and can be controlled by the targeted mode, ${{\bf{a}}_{\bf{t}}}$. (b) The states of the WPT system are controlled by the coupling coefficients ${\kappa _{1m}}$ and ${\kappa _{2m}}$. PT-symmetric states can form on the dashed line where $\left| {{\kappa _{1m}}} \right| = \left| {{\kappa _{2m}}} \right|$. $\omega_i$ is shown as the 3D plots, where the x and y coordinates represent $\Delta {\kappa _{1m}}$ and $\Delta {\kappa _{1m}}$ deviate away from the dashed line, showing that the PT-symmetric state is less stable with a lower overall coupling in the strong coupling regime and becomes unstable in the weak coupling regime.}
\end{figure}

We tailor the metamaterial’s resonant mode, which ultimately controls the ratio between ${\kappa _{1m}}$ and ${\kappa _{2m}}$. As illustrated in FIG. 1(b), the states of the system depend on ${\kappa _{1m}}$ and ${\kappa _{2m}}$. Notably, a PT-symmetric state, characterized by the absence of frequency splitting, emerges when the coupling coefficients are balanced, $\left| {{\kappa _{1m}}} \right| = \left| {{\kappa _{2m}}} \right|$. This state remains stable in the strong coupling regime, where its the stability increases in proportion to the overall coupling strength, ${\kappa _{1m}}^2 + {\kappa _{2m}}^2$; conversely, it becomes unstable in the weak coupling regime.

The metamaterial-enhanced WPT system can be modeled using the time-independent coupled-mode theory:
\begin{equation}
\frac{d}{{dt}}\left[ {\begin{array}{*{20}{c}}
{{a_1}}\\
{{{\bf{a}}_{\bf{m}}}}\\
{{a_2}}
\end{array}} \right] = \left[ {\begin{array}{*{20}{c}}
{i{\omega _1} + {g_1}}&{ - i{{\bf{\kappa }}_{1{\bf{u}}}}}&{ - i{\kappa _0}}\\
{ - i{{\bf{\kappa }}_{1{\bf{u}}}}}&{{{\bf{H}}_{\bf{m}}} - {\gamma _u}}&{ - i{{\bf{\kappa }}_{{\bf{2u}}}}}\\
{ - i{\kappa _0}}&{ - i{{\bf{\kappa }}_{{\bf{2u}}}}}&{i{\omega _2} - {\gamma _2}}
\end{array}} \right]\left[ {\begin{array}{*{20}{c}}
{{a_1}}\\
{{{\bf{a}}_{\bf{m}}}}\\
{{a_2}}
\end{array}} \right],
\end{equation}
where $a_1$ and $a_2$ are resonance amplitudes of the Tx and Rx resonators, ${{\bf{a}}_{\bf{m}}} = {\left[ {\begin{array}{*{20}{c}}
{{a_{u1}}}& \cdots &{{a_{uN}}}
\end{array}} \right]^T}$, represents the metamaterial’s mode. $\omega_1$ and $\omega_2$ are the resonance frequencies of the Tx and Rx resonators. We choose ${\omega _1} = {\omega _2} = {\omega _0}$ with $\omega_0$ being the operating frequency. $g_1$ is the net gain of the Tx resonator that takes into account its intrinsic loss, ${g_1} = {g_{10}} - {\gamma _{10}}$, ${g_{10}}$ is the input gain, and $\gamma_{10}$ is the intrinsic loss of the Tx resonator. $\gamma_u$ is damping coefficient of the unit cell, assumed to be identical across the metamaterial; and   is the damping of the Rx resonator, respectively. ${{\bf{H}}_{\bf{m}}}$ is the lossless Hamiltonian of the metamaterial, ${{\bf{H}}_{\bf{m}}} = \left[ {\begin{array}{*{20}{c}}
{i{\omega _{u1}}}& \ldots &{ - i{\kappa _{u1N}}}\\
 \vdots & \ddots & \vdots \\
{ - i{\kappa _{uN1}}}& \ldots &{i{\omega _{uN}}}
\end{array}} \right]$, where $\kappa_{uij}$ is the coupling coefficient between the metamaterial’s i-th and j-th unit cells, and   is the resonance frequency of the j-th unit cell. The metamaterial configuration is determined by the distribution of the unit cells’ resonance frequencies, $\omega_{uj}$ is the input gain, and $\omega_{uj}$ is the intrinsic loss of the Tx resonator. For simplicity in deriving the solution, we assume the coupling coefficients to be non-dispersive. This assumption is largely accurate and the resulting error becomes negligible particularly when the system’s eigenfrequency approaches the operating frequency, $\omega_{0}$.

We define the targeted mode of the metamaterial as ${{\bf{a}}_{\bf{t}}} = {\left[ {\begin{array}{*{20}{c}}
{{a_{t1}}}& \cdots &{{a_{tN}}}
\end{array}} \right]^T}$. To simply the calculation we normalize the overall intensity of this mode, ${{\bf{a}}_{\bf{t}}}^T{{\bf{a}}_{\bf{t}}} = 1$. The metamaterial’s configuration is controlled by adjusting the resonance frequencies of its unit cells, forming the targeted mode at the operating frequency without perturbation from the Tx and Rx resonators, i.e., $i{\omega _0}{{\bf{a}}_{\bf{m}}} = {{\bf{H}}_{\bf{m}}}{{\bf{a}}_{\bf{m}}}$. As a result, we can solve for the metamaterial’s configuration as
\begin{equation}
{\omega _{uj}} = {\omega _0} + \sum\limits_{i = 1}^{N,i \ne j} {{\kappa _{uij}}{a_{ti}}} /{a_{tj}}.
\end{equation}

As outlined in the perturbation theory (Supplemental Material, Sec. 1 \cite{wangSI2023}), this system can be simplified into a three-body system, in which the metamaterial is treated as a single resonator,
\begin{equation}
i\omega {\bf{a}} = {{\bf{H}}_{{\bf{3b}}}}{\bf{a}},
\end{equation}
where ${\bf{a}} = {\left[ {\begin{array}{*{20}{c}}
{{a_1}}&{{a_m}}&{{a_2}}
\end{array}} \right]^T}$
and ${{\bf{H}}_{{\bf{3b}}}} = \left[ {\begin{array}{*{20}{c}}
{i{\omega _0} + {g_1}}&{ - i{\kappa _{1m}}}&{ - i{\kappa _0}}\\
{ - i{\kappa _{1m}}}&{i{\omega _0} - {\gamma _u}}&{ - i{\kappa _{2m}}}\\
{ - i{\kappa _0}}&{ - i{\kappa _{2m}}}&{i{\omega _0} - {\gamma _2}}
\end{array}} \right]$.
Note that the inverse design of the metamaterial is only valid at $\omega_0$; therefore, the reduction from the complex many-body system, as described by Eq. (1), to the more simplified three-body system, as described by Eq. (3), is only meaningful and accurate when the system’s real eigenfrequency is in close proximity to $\omega_0$.

The eigenstate corresponding to Eq. (3) follows the time evolution of ${e^{i\omega t}}$. The characteristic equation of the three-body system is $\det ({{\bf{H}}_{{\bf{3b}}}} - i\omega {\bf{I}}) = 0$, where $\omega$ is the eigenfrequency. Solving this characteristic equation can be quite complex. However, the complexity can be significantly reduced under certain conditions: when the original coupling between the Tx and Rx resonators are weak, ${\kappa _0} \approx 0$; when the metamaterial exhibits low loss ${\gamma _m} \approx 0$, a condition required for the system to maintain PT symmetry (Supplemental Material, Sec. 2 \cite{wangSI2023}); and when the net gain of the Tx resonator counterbalances the net loss of the Rx resonator, ${g_1} = {\gamma _2}$. When these conditions are met, the characteristic equation can be simplified to 
\begin{equation}
\Delta \omega \,\left[ {\Delta {\omega ^2} - \left( {s - {\gamma _2}^2} \right)} \right]\, = i\delta {\gamma _2},
\end{equation}
where $\Delta \omega  = \omega  - {\omega _0}$, $s = {\kappa _{2m}}^2 + {\kappa _{1m}}^2$, represents the total coupling strength bridged by the metamaterial; $\delta  = {\kappa _{2m}}^2 - {\kappa _{1m}}^2$, represents the difference between the metamaterial’s couplings to Rx and Tx.

When the coupling coefficients are perfectly matched, i.e., ${\kappa _{2m}}^2 = {\kappa _{1m}}^2$, the characteristic equation becomes $\Delta \omega \,\left[ {\Delta {\omega ^2} - \left( {s - {\gamma _2}^2} \right)} \right]\, = 0$, which yields a state that 
\begin{equation}
\Delta \omega  = 0.
\end{equation}

The eigenfrequency is real, representing PT-symmetric states. Please note that while two additional states are indicated by $\Delta \omega  =  \pm \sqrt {s - {\gamma _2}^2} $, the metamaterial’s design, as per Eq. (2), is precisely designed for $\omega_0$, making these additional states potentially unobservable in practical scenarios. 

Eq. (5) suggests to tune the targeted mode of the metamaterial so that $\delta  = 0$ is achieved. Theoretically, this state can exist under any overall coupling strength. However, it will be considered highly unstable if this state can only exist at the perfectly matched condition of $\delta  = 0$. Such a precise requirement is undesirable because even a minor mismatch between ${\kappa _{2m}}^2$ and ${\kappa _{1m}}^2$ will lead to annihilation of this state.

To analyze the stability of this state, we focus on the scenarios where the coupling coefficients are unbalanced, meaning ${\kappa _{2m}}^2 \ne {\kappa _{1m}}^2$. In these cases of unbalanced coupling, the characteristic equation, as presented in Eq. (4), yields
\begin{equation}
    \Delta \omega  = {\omega _i}i,
\end{equation}
where $\omega_i$ is the imaginary component of the eigenfrequency. While Eq. (4) also gives rise to other solutions with ${\omega _r} \ne {\omega _0}$, these are likely unobservable in practical conditions due to the metamaterial being inversely designed specifically for $\omega_0$. Incorporating Eq. (6) into Eq. (4) results in
\begin{equation}
{\omega _i}^2 + (s - {\gamma _2}^2) =  - \delta {\gamma _2}/{\omega _i}
\end{equation}

\begin{figure}[b]
\includegraphics[width=0.47\textwidth]{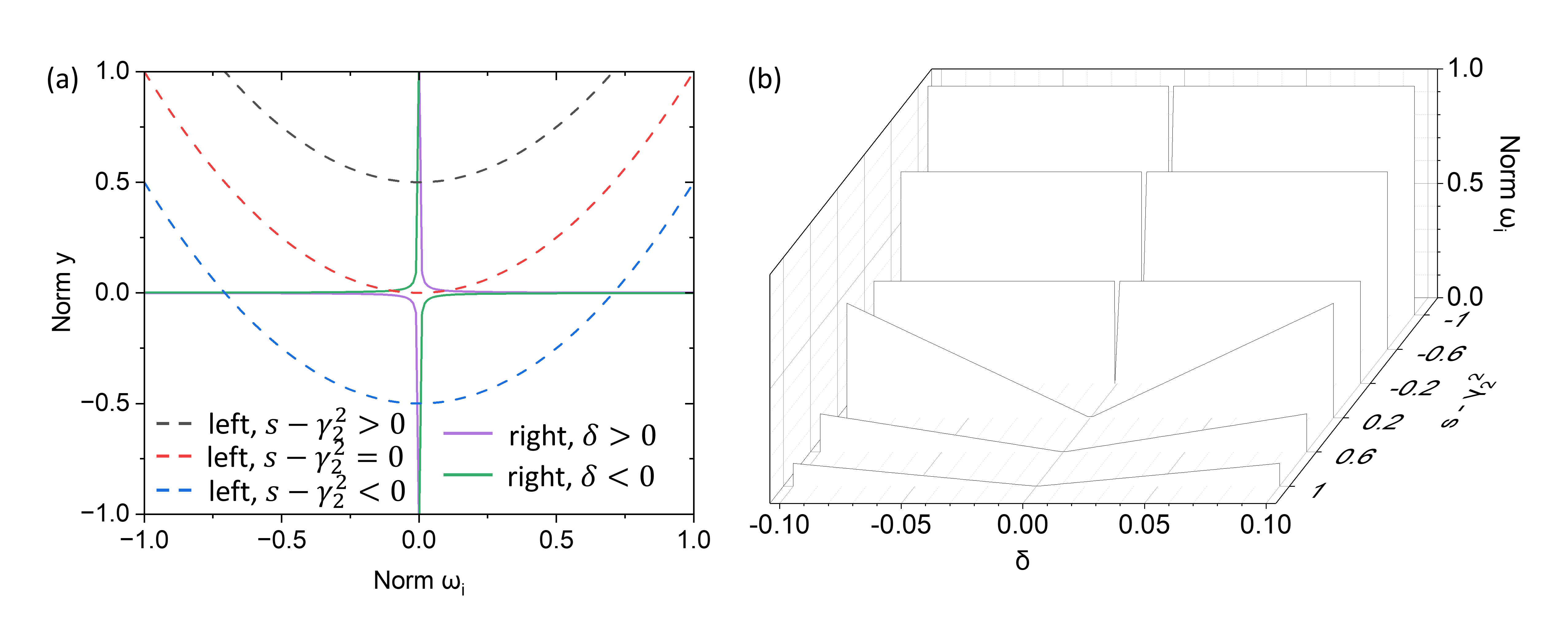}
\caption{System states at the mismatched coupling condition, ${\kappa _{2m}}^2 \ne {\kappa _{1m}}^2$. (a) plots of the two sides of Eq. (7), $y({\omega _i}) = {\omega _i}^2 + (s - {\gamma _2}^2)$, $y({\omega _i}) =  - \delta {\gamma _2}/{\omega _i}$, as dashed and solid curves, respectively. $\omega_i$ and y are both normalized to the range of -1 to 1, denoted by Norm y and Norm $\omega_i$. (b) $\omega_i$ versus $\delta$ at different overall coupling strength, characterized by $s - {\gamma _2}^2$. ${\gamma _2}$ is set as one here for demonstrating of the concept.}
\end{figure}

\begin{figure}[b]
\includegraphics[width=0.47\textwidth]{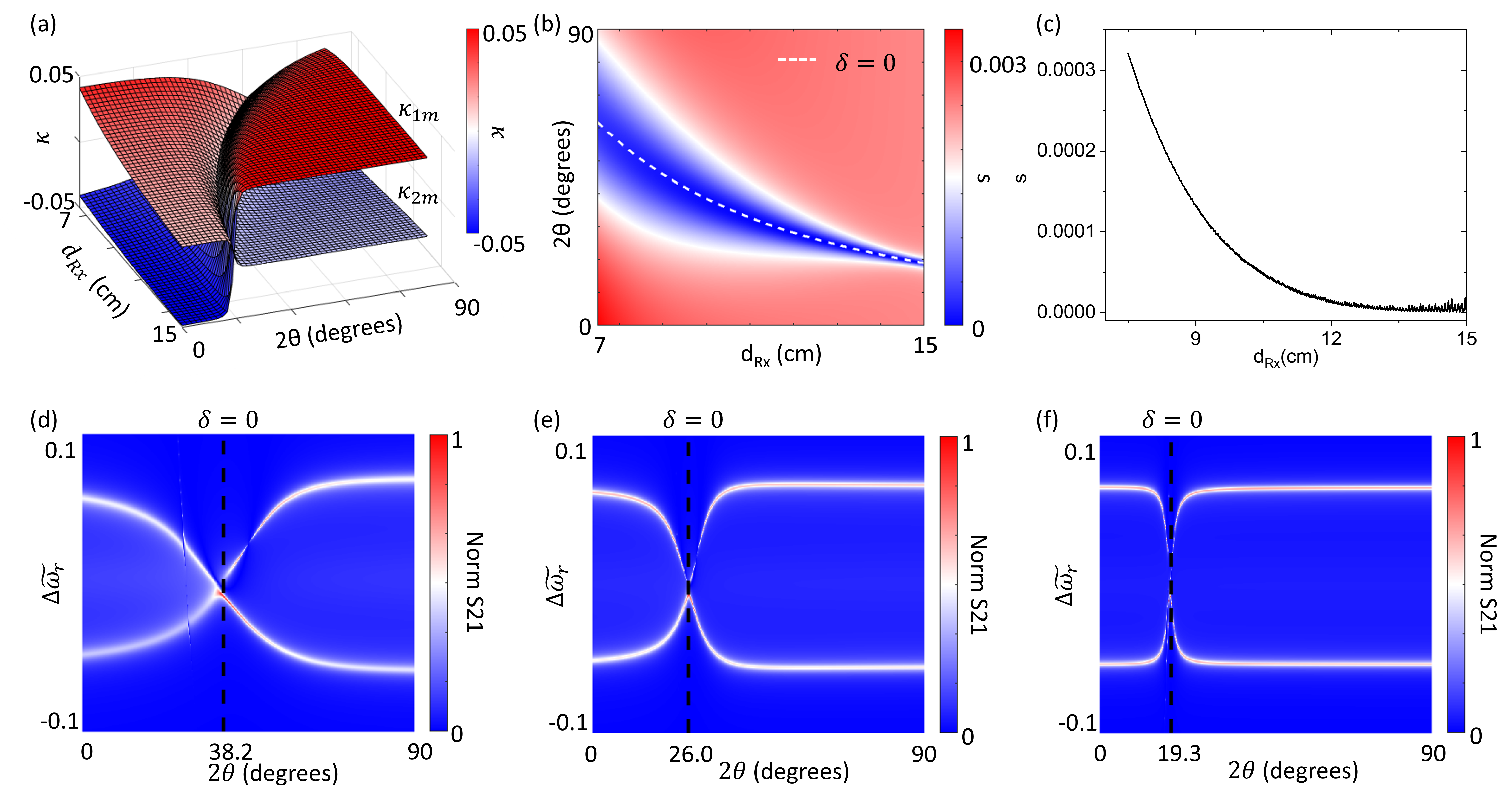}
\caption{Practical system with the coupling coefficient controlled by $\theta$. (a) $\kappa_{1m}$ and $\kappa_{2m}$ as a function of $\theta$ and the distance between the metamaterial and Rx, $d_{Rx}$. (b) Total coupling strength, $s$, as a function of $\theta$ and $\delta$ with the position of the $\delta = 0$ condition. (c) Total coupling strength, $s$, at the critical $\theta$ versus $d_{Rx}$. (d-f) Numerically calculated scattering parameter, $S_{21}$, of the system. The metamaterial-Tx distance is fixed at 10 cm, and the metamaterial-Rx distance varies at (d) 10 cm, (e) 12.5 cm, and (f) 15 cm. The dashed line indicates where $\left| {{\kappa _{1m}}} \right| = \left| {{\kappa _{2m}}} \right|$. The load resistance is $10 \Omega$.}
\end{figure}

Although solving Eq. (7) analytically poses challenges, it can be analyzed graphically. As shown in FIG. 2(a), the left side of the equation is a parabola as a function of $\omega_i$, and the right side is a hyperbola versus $\omega_i$. The intersection of the two curves provides the solution. When $\omega_i$ is non-zero, indicating the eigenfrequeny is complex, and the state is thus anti-PT-symmetric. To investigate the phase transition around $\delta  = 0$, we examine the scenarior where $\delta  \to 0$. In this condition, the hyperbolic terms on the right side of Eq. (7) approach towards the horizontal and vertical axes. If $s - {\gamma _2}^2 > 0$, we can simplify the equation $ - \delta {\gamma _2}/{\omega _i} \approx (s - {\gamma _2}^2)$; otherwise, the equation is approximated by ${\omega _i}^2 + (s - {\gamma _2}^2) \approx 0$. The time evolution of the eigenstate here is represented by ${e^{i{\omega _0}t}}{e^{ - {\omega _i}t}}$. For a state to be stable, the energy needs to decay over time, therefore, $\omega_i$ must be greater than or equal to zero. As a result, the conditions for stable states are
\begin{equation}
\Delta \omega (\delta  \to 0) = \left\{ {\begin{array}{*{20}{c}}
{i\frac{{\left| \delta  \right|{\gamma _2}}}{{s - {\gamma _2}^2}}}&{s - {\gamma _2}^2 \ge 0}\\
{i\sqrt {{\gamma _2}^2 - s} \,}&{otherwise}
\end{array}} \right.,
\end{equation}
where the condition $s - {\gamma _2}^2 \ge 0$ denotes the strong coupling regime, and the rest represents the weak coupling regime. The corresponding distribution is illustrated in FIG. 2(b). Notably, the transition at $\delta  = 0$ is not continous, characterizing it as a first order phase transition. As any coupling mismatch will lead to a significant decay of the state, represented by $\omega_i$, the PT-symmetric state at $\delta = 0$ lacks stability. In the strong coupling regime, $\Delta \omega  = i\frac{{\left| \delta  \right|{\gamma _2}}}{{s - {\gamma _2}^2}}$ trends towards 0 as $\delta  \to 0$, representing a second order phase transition. The state remains approximately PT-symmetric even with a minor mismatch between the coupling intensities, as given by $\kappa_{1m}^2$ and $\kappa_{2m}^2$. As a result, the PT-symmetric state at $\delta = 0$ is stable. 

To control $\kappa_{1m}$ and $\kappa_{2m}$, we design ${{\bf{a}}_{\bf{t}}}$ as a composite of two distinct modes, each targeting the positions of the Tx and Rx resonators, defined as ${{\bf{a}}_{{\bf{t - Tx}}}}$ and ${{\bf{a}}_{{\bf{t - Rx}}}}$, respectively. To regulate the two modes, we adjust the ratios between $\kappa_{1m}^2$ and $\kappa_{2m}^2$ through a parameter $\theta$. We define $\theta$ with a range from 0 to 90 degrees, allowing us to continously  adjust the ratio from zero to infinity and vary the composition of ${{\bf{a}}_{{\bf{t - Tx}}}}$ and ${{\bf{a}}_{{\bf{t - Rx}}}}$. Here, $\theta$ does not represent a physical angle but is rather utilized to tune the intensity ratios of the modes. This, in turn,  enables precise control over the coupling coefficients $\kappa_{1m}$ and $\kappa_{2m}$.
\begin{equation}
{{\bf{a}}_{\bf{t}}} \propto \sin \theta {{\bf{a}}_{{\bf{t - Tx}}}} + \cos \theta {{\bf{a}}_{{\bf{t - Rx}}}}
\end{equation}
While in theory, ${{\bf{a}}_{{\bf{t - Tx}}}}$ and ${{\bf{a}}_{{\bf{t - Rx}}}}$ can be selected from a range of distributions, we optimize them to mirror the coupling coefficient distribution for minimizing the loss of the metamaterial. We set ${{\bf{a}}_{{\bf{t - Tx}}}} = {{\bf{\kappa }}_{{\bf{1u}}}}$ and ${{\bf{a}}_{{\bf{t - Rx}}}}{\bf{ = }}{{\bf{\kappa }}_{{\bf{2u}}}}$ (Supplemental Material, Sec. 3 \cite{wangSI2023}).

To practically demonstrate our theory, we use a Tx coil with a 10 cm radius and a Rx coil with a 5 cm radius. Each coil contains  a single turn to minimize perturbations to the metamaterial’s mode. The metamaterial itself consists of a 3-by-3 square array of unit cells, with a periodicity of 10 cm in both the x and y directions. Each unit cell, having a 4.55 cm radius  and 5 turns, is designed to form resonance in the tens of megahertz range. This design ensures strong near-field coupling among adjacent unit cells. The unit cells are arranged in a non-intersecting planar array. 

As shown in FIG. 3(a), by controlling $\theta$, we can regulate the amplitude of $\kappa_{1m}$ and $\kappa_{2m}$ to compensate change induced by different $d_{Rx}$. We provide the specific details of the metamaterial configuration in the Supplemental Material, Sec. 3 \cite{wangSI2023}. Specifically, when $\theta$ is near 0 degrees, the coupling of the metamaterial is more oriented towards the Tx. Conversely, when $\theta$ is near 90 degrees, the coupling shifts more towards the Rx. A phase transition occurs when the couplings are balanced, i.e., $\delta  = 0$, as indicated by the dashed lines in FIG. 3(b). As Rx moves further away from the metamaterial, meaning $d_{Rx}$ increaes, the metamaterial requires a stronger mode relative to the Rx (${{\bf{a}}_{{\bf{t - Rx}}}}$) to maitain balanced coupling. Consequently, the phase transition point shifts to a lower $\theta$. FIG. 3(c) shows that the total coupling strength, $s$, decreases as $d_{Rx}$ increases, leading to increased instability of the PT-symmetric state and a more rapid phase transition. 

To probe the resonance states of the system, we measured the scattering parameter, $S_{21}$, between the Tx and Rx resonators. As detailed in the Supplemental Material, Sec. 4, we have established a correlation between $S_{21}$ under an oscillating voltage input and the PT-symmetric states of the system \cite{wangSI2023}.  FIG. 3(d) – 3(f)) show our simulation results for $S_{21}$ as a function of distances, $d_{Rx}$, of 10 cm, 12.5 cm, and 15 cm between the metamaterial and the Rx, while maintaining a constant distance of 10 cm between the metamaterial and the Tx. These figures validate our theory, accurately pinpointing the position of the PT-symmetric states (with $\delta = 0$), as marked by the dashed lines. The theory also precisely predicts the movement and stability of the PT-symmetric states as the Rx position changes. Specifically, an increase in the distance between the metamaterial and the Rx resonator requires a higher ${{\bf{a}}_{{\bf{t - Rx}}}}$ to achieve balanced couplings, which consequently causes the exceptional point to shift towards a lower $\theta$ value. Furthermore, with the decrease in the overall coupling  , the PT-symmetric state becomes more unstable. As a result, as $d_{Rx}$ increases, $\theta$, at which the frequency splitting of the resonance states begins to shift, decreases, leading to a more rapid phase transition as depicted in FIG. 3(f). If $d_{Rx}$ continues to increase, the PT-symmetric state eventually reaches a level of instability where it cannot practically exsit.

While our theory offers insights, it is crucial to acknowledge its limitations. Accurate predictions are contingent on the following conditions: (1) the frequency must be set at $\omega_0$, (2) the imbalance $\delta$ should be relatively small, and (3) the coupling between the Tx and Rx is assumed to be weak. The dispersive states illustrated in FIG. 3(d) – 3(f) cannot be quantitively predicted with the abovementioned theory.  However, we can qualitatively interpret these states.

The metamaterial plays a pivotal role by enabling the adjustment of $\theta$ to its critical value, corresponding to the balanced coupling strength where ${\kappa _{1m}}^2 = {\kappa _{2m}}^2$. These states are shown in the numerical examples of FIG. 3(d)-3(f), where the critical $\theta$ values are $38.2^{\circ}$, $26^{\circ}$, and $19.3^{\circ}$, respectively. In each scenario, when $\theta$ deviates from its critical value, the metamaterial cannot effectively bridge the Tx and Rx resonators; consequently, the system reverts to a behavior similar to a two-body system, characterized by the two frequency splitting states. In contrast, as $\theta$ converges to the critical value, a phase transition to a state devoid of frequency splitting is observed. In line with our previously outlined theory, this state is PT-symmetric, leading to an increased $S_{21}$. The phase transition point in this multi-body system, shifting from the PT-symmetric and anti-PT-symmetric states, is identified as a high-order exceptional point \cite{hao2023frequency}. Additionally, the discontinuities to the left of the critical $\theta$ in FIG 3(d) are associated with conditions where ${\kappa _{1m}} =  - {\kappa _{2m}}$, placing the system in the weak coupling regime. Here, the PT-symmetric state is unstable, and the (first order) transition occurs very rapidly.

To validate our theory, we established an experimental setup comprising a Tx resonator, a metamaterial with 9 elements, and an Rx resonator, as shown in FIG. 4(a). The metamaterial’s unit cells were designed as open-ended spirals with varying radii: the middle loop has a smaller radius of 2.65 cm, while the top and bottom loops have a larger radius of 4.55 cm. This design facilitates a relatively uniform magnetic field in the vertical direction and offers tunability when compressing the structure (FIG. 4(b)). The detailed simulation model is available in the Supplemental Materials, Sec. 5 \cite{wangSI2023}. We observed that the measured resonance frequencies as a function of the unit cell’s height align well with our simulation results. In replicating the simulation, we used a voltage source to drive the Tx resonator and measured $S_{21}$ of the system (see Supplemental Material, Sec. 6 for the measurement setup \cite{wangSI2023}). The resonance frequencies of the unit cells, controlled by the heights of the spiral resonators, are demonstrated in FIG. 4(b). As shown in FIG. 4(c), the scattering parameter exhibits stronger intensity at $\theta \approx 35 ^\circ$ , suggesting a PT-symmetric state. We also observed a phase transition from two anti-PT-symmetric states to a single PT-symmetric state. Reflecting the numerical results shown in FIG. 3, the anti-PT-symmetric states flanking the critical condition exhibit asymmetry.

We further conducted experiments to compensate for the changes in the separation between the metamaterial and the Rx resonator. FIG. 4(d) shows that increasing this separation causes a leftward shift of the critical $\theta$, consistent with our theoretical prediction. However, a minor discrepancy between the experiment and the theory was noted. This discrepancy arises because the separation between the metamaterial and the Rx resonator was overestimated. Given that different unit cells have varied heights, accurately determining the effective z-position of the metamaterial is challenging. Therefore, we used the bottom of the metamaterial as a reference for determining $d_{Rx}$. This approach introduced some errors in calculating $\theta$, especially when the Rx resonator is positioned close to the metamaterial.

\begin{figure}
\includegraphics[width=0.5\textwidth]{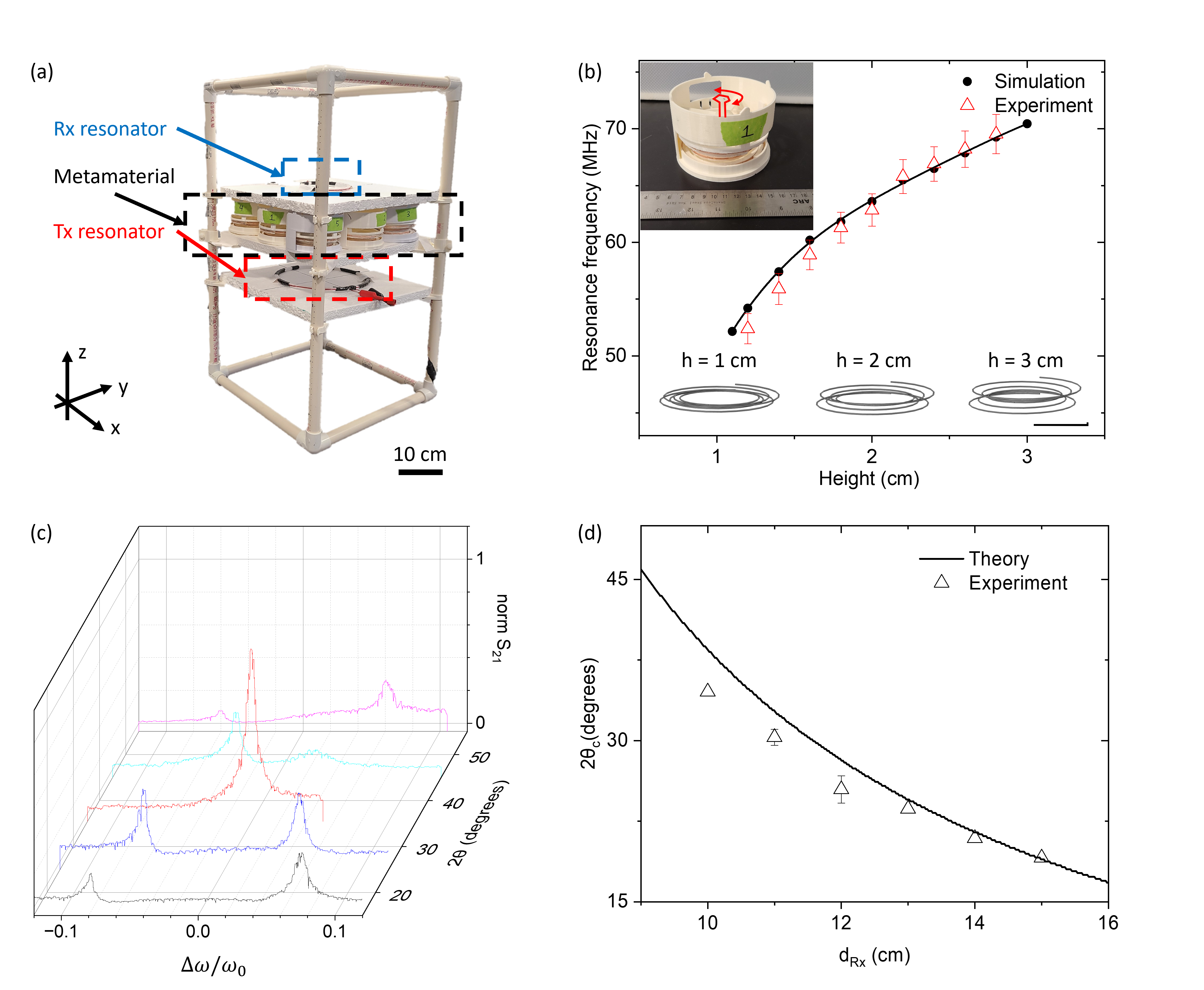}
\caption{Measurement of the eigenstates. (a) Experimental setup. (b) Measured and simulated resonance frequency versus height of the resonator. The error bars represent standard deviation of the 9 unit cells. The inset picture shows the unit cell with a tunable height by the top screw. (c) Measured spectrum of the normalized $S_{21}$ at different configurations. (d) $2\theta$ for the critical condition, $2\theta_c$, versus the distance between the metamaterial and the Rx resonator, $d_{Rx}$. The error bars represent standard deviation of three measurements.}
\end{figure}

In summary, we present a method to control PT symmetry in non-Hermitian WPT systems using a metamaterial, which functions as a controllable relay for tuning coupling coefficients to the Tx and Rx resonators. Achieving a balance between $\kappa_{1m}$ and $\kappa_{2m}$ allows for PT symmetry, even with asymmetrically positioned and sized Tx and Rx resonators. Validated by simulation and experiments, our results align with theoretical predictions and establish a foundation for metamaterial-controlled PT symmetry in near-field applications, opening avenues for developing high-efficiency, free-positioning WPT systems.

\bibliography{apssamp}

\end{document}


\preprint{APS/123-QED}

\title{Supplemental Material to ``Metamaterial-Controlled Parity-Time Symmetry in Non-Hermitian Wireless Power Transfer Systems"}

\author{Hanwei Wang}
\affiliation{%
 Department of Electrical and Computer Engineering, University of Illinois at Urbana-Champaign, Urbana, IL, US.
}%
 \altaffiliation[Also at ]{Holonyak Micro and Nanotechnology Laboratory, University of Illinois at Urbana-Champaign, Urbana, IL, US.}

 \author{Joshua Yu}%
\affiliation{%
 Department of Electrical and Computer Engineering, University of Illinois at Urbana-Champaign, Urbana, IL, US.
}

\author{Xiaodong Ye}%
\affiliation{%
 Department of Electrical and Computer Engineering, University of Illinois at Urbana-Champaign, Urbana, IL, US.
}%
 \altaffiliation[Also at ]{Holonyak Micro and Nanotechnology Laboratory, University of Illinois at Urbana-Champaign, Urbana, IL, US.}

\author{Yang Zhao}%
 \email{yzhaoui@illinois.edu}
\affiliation{%
 Department of Electrical and Computer Engineering, University of Illinois at Urbana-Champaign, Urbana, IL, US.
}%
 \altaffiliation[Also at ]{Holonyak Micro and Nanotechnology Laboratory, University of Illinois at Urbana-Champaign, Urbana, IL, US.}

\date{\today}

\maketitle
\onecolumngrid

\subsection*{\Large Section 1. Simplification to the three-body system}
We apply perturbation theory to Eq. (1) in the main text
\begin{equation}
i\omega {\left[ {\begin{array}{*{20}{c}}
{{a_1}}\\
{{{\bf{a}}_{\bf{m}}}}\\
{{a_2}}
\end{array}} \right]^\dag }\left[ {\begin{array}{*{20}{c}}
{{a_1}}\\
{{{\bf{a}}_{\bf{m}}}}\\
{{a_2}}
\end{array}} \right] = {\left[ {\begin{array}{*{20}{c}}
{{a_1}}\\
{{{\bf{a}}_{\bf{m}}}}\\
{{a_2}}
\end{array}} \right]^\dag }\left[ {\begin{array}{*{20}{c}}
{i{\omega _1} + {g_1}}&{ - i{{\bf{\kappa }}_{1{\bf{u}}}}}&{ - i{\kappa _0}}\\
{ - i{{\bf{\kappa }}_{1{\bf{u}}}}}&{{{\bf{H}}_{\bf{m}}} - {\gamma _u}}&{ - i{{\bf{\kappa }}_{{\bf{2u}}}}}\\
{ - i{\kappa _0}}&{ - i{{\bf{\kappa }}_{{\bf{2u}}}}}&{i{\omega _2} - {\gamma _2}}
\end{array}} \right]\left[ {\begin{array}{*{20}{c}}
{{a_1}}\\
{{{\bf{a}}_{\bf{m}}}}\\
{{a_2}}
\end{array}} \right].
\tag{A1}
\end{equation}
After inverse design through Eq. (2) in the main text, the metamaterial’s Hamiltonian follows 
$i{\omega _0}{{\bf{a}}_{\bf{m}}} = {{\bf{H}}_{\bf{m}}}{{\bf{a}}_{\bf{m}}}$. Therefore, Eq. (A1) could be simplified as
\begin{equation}
i\omega {{\bf{a}}^{\bf{\dag }}}{\bf{a}} = {{\bf{a}}^{\bf{\dag }}}{{\bf{H}}_{{\bf{3b}}}}{\bf{a}}.
\tag{A2}
\end{equation}
where the system state ${\bf{a}} = {\left[ {\begin{array}{*{20}{c}}
{{a_1}}&{{a_m}}&{{a_2}}
\end{array}} \right]^T}$, three-body Hamiltonian ${{\bf{H}}_{{\bf{3b}}}} = \left[ {\begin{array}{*{20}{c}}
{i{\omega _0} + {g_1}}&{ - i{\kappa _{1m}}}&{ - i{\kappa _0}}\\
{ - i{\kappa _{1m}}}&{i{\omega _0} - {\gamma _u}}&{ - i{\kappa _{2m}}}\\
{ - i{\kappa _0}}&{ - i{\kappa _{2m}}}&{i{\omega _0} - {\gamma _2}}
\end{array}} \right]$. It is easy to prove that $i\omega {\bf{a}} = {{\bf{H}}_{{\bf{3b}}}}{\bf{a}}$, as Eq. (3) in the main text, is a sufficient condition to Eq. (A2). When the state ${\bf{a}}$ is PT-symmetric, we can show that Eq. (3) is the sufficient and necessary condition using proof by contradiction. We first assume there is a state ${\bf{b}}$, being different from ${\bf{a}}$, that is yielded by the three-body Hamiltonian 
\begin{equation}
i\omega {\bf{b}} = {{\bf{H}}_{{\bf{3b}}}}{\bf{a}}.
\tag{A3}
\end{equation}
We set the difference between the two states as ${\bf{\varsigma }}$, ${\bf{\varsigma }} = {\bf{b}} - {\bf{a}}$. Eq. (A3) yields
\begin{equation}
{{\bf{\varsigma }}^\dag }{\bf{\varsigma }} = \frac{1}{{{\omega ^2}}}\left( {{{\bf{a}}^\dag }{{\bf{H}}_{{\bf{3b}}}}^H + i\omega {{\bf{a}}^\dag }} \right)\left( {{{\bf{H}}_{{\bf{3b}}}}{\bf{a}} - i\omega {\bf{a}}} \right).
\tag{A4}
\end{equation}
Eq. (A4) can be simplified as ${{\bf{\varsigma }}^\dag }{\bf{\varsigma }} = \frac{1}{{{\omega ^2}}}\left( {{{\bf{a}}^\dag }{{\bf{H}}_{{\bf{3b}}}}^H{{\bf{H}}_{{\bf{3b}}}}{\bf{a}} + i\omega {{\bf{a}}^\dag }{{\bf{H}}_{{\bf{3b}}}}{\bf{a}} - i\omega {{\bf{a}}^\dag }{{\bf{H}}_{{\bf{3b}}}}^H{\bf{a}} + {\omega ^2}{{\bf{a}}^\dag }{\bf{a}}} \right)$. According to Eq. (A2), ${{\bf{a}}^\dag }{{\bf{H}}_{{\bf{3b}}}}{\bf{a}} = i\omega {{\bf{a}}^{\bf{\dag }}}{\bf{a}}$, ${{\bf{a}}^\dag }{{\bf{H}}_{{\bf{3b}}}}^H{\bf{a}} =  - i\omega {{\bf{a}}^{\bf{\dag }}}{\bf{a}}$. Furthermore, for PT-symmetric state, the dyad tensor 
\begin{equation}
{\bf{a}}{{\bf{a}}^\dag } = {\bf{I}}.
\tag{A5}
\end{equation}
Therefore ${{\bf{\varsigma }}^\dag }{\bf{\varsigma }} = \frac{1}{{{\omega ^2}}}\left( {{{\bf{a}}^\dag }{{\bf{H}}_{{\bf{3b}}}}^H{{\bf{H}}_{{\bf{3b}}}}{\bf{a}} - {\omega ^2}{{\bf{a}}^\dag }{\bf{a}}} \right) = \frac{1}{{{\omega ^2}}}\left( {{{\bf{a}}^\dag }{{\bf{H}}_{{\bf{3b}}}}^H{\bf{a}}{{\bf{a}}^\dag }{{\bf{H}}_{{\bf{3b}}}}{\bf{a}} - {\omega ^2}{{\bf{a}}^\dag }{\bf{a}}} \right) = 0$. As the difference between ${\bf{b}}$ and ${\bf{a}}$ has a zero magnitude, ${\bf{b}} = {\bf{a}}$. This proof means that the PT-symmetric state given by the three-body equation is also unique to the multibody equation.

\subsection*{\Large Section 2. \uppercase{PT}-symmetry of the Hamitonian}
PT-symmetric Hamiltonians follow 
\begin{equation}
\hat P\hat T({\bf{H}}){\hat T^{ - 1}}{\hat P^{ - 1}} = {\bf{H}},
\tag{A6}
\end{equation}
where $\hat P$ is parity operator, and $\hat T$ is time-reversal operator. For the three-body system in the frequency domain, $\hat P = \left[ {\begin{array}{*{20}{c}}
{}&{}&1\\
{}&1&{}\\
1&{}&{}
\end{array}} \right]$, $\hat T(i){\hat T^{ - 1}} =  - i$. We can see that for the three-body Hamiltonian, ${{\bf{H}}_{{\bf{3b}}}}$,
\begin{equation}
\hat P\hat T({{\bf{H}}_{{\bf{3b}}}}){\hat T^{ - 1}}{\hat P^{ - 1}} = \left[ {\begin{array}{*{20}{c}}
{i{\omega _0} - {g_1}}&{ - i{\kappa _{1m}}}&{ - i{\kappa _0}}\\
{ - i{\kappa _{1m}}}&{i{\omega _0} + {\gamma _u}}&{ - i{\kappa _{2m}}}\\
{ - i{\kappa _0}}&{ - i{\kappa _{2m}}}&{i{\omega _0} + {\gamma _2}}
\end{array}} \right],
\tag{A7}
\end{equation}
where $g = \gamma  + {\gamma _m}$. The system is PT-symmetric, $\hat P\hat T({{\bf{H}}_{{\bf{3b}}}}){\hat T^{ - 1}}{\hat P^{ - 1}} = {{\bf{H}}_{{\bf{3b}}}}$, only when ${\gamma _m} = 0$.

\subsection*{\Large Section 3. Metamaterial configuration}
The targeted modes of the metamateial, ${{\bf{a}}_{{\bf{t - Tx}}}}$ and ${{\bf{a}}_{{\bf{t - Rx}}}}$, can theortically be chosen arbitrarily if the metamaterial has a zero loss. However, in practice, the metamaterial is not loss-free. We would want to miniaturize the potential loss caused by the resistive loss of the metamaterial resonators. To achive this, we use the following optimization considering the metamaterial’s loss.

The efficiency is given by
\begin{equation}
\eta  = \frac{{2{\gamma _L}{{\left| {{a_2}} \right|}^2}}}{{2{\gamma _{10}}{{\left| {{a_1}} \right|}^2} + 2{\gamma _m}{{\left| {{{\bf{a}}_{\bf{m}}}} \right|}^2} + 2({\gamma _{20}} + {\gamma _L}){{\left| {{a_2}} \right|}^2}}}.
\tag{A8}
\end{equation}
The current distribution of the metamaterial is proportional to the normalized targeted mode, ${{\bf{a}}_t}$, as ${{\bf{a}}_{\bf{m}}} = {a_m}{{\bf{a}}_t}$. At the operating frequency, ${\omega _0}$, Eq. (3) of the main text yields that
\begin{equation}
{a_m} = \frac{{ - (i{{\bf{\kappa }}_{{\bf{1u}}}}^{\bf{\dag }}{{\bf{a}}_{\bf{t}}}{a_1} + i{{\bf{\kappa }}_{{\bf{2u}}}}^{\bf{\dag }}{{\bf{a}}_{\bf{t}}}{a_2})}}{{{\gamma _m}}},
\tag{A9}
\end{equation}
and
\begin{equation}
{a_2} = \frac{{ - (i{\kappa _0}{a_1} + i{{\bf{\kappa }}_{{\bf{2u}}}}^{\bf{\dag }}{{\bf{a}}_{\bf{t}}}{a_m})}}{{{\gamma _2}}}.
\tag{A10}
\end{equation}
Combining equations (A9) and (A10), we get
\begin{equation}
{a_2} =  - \frac{{i{\kappa _0}{a_1}{\gamma _m} + ({{\bf{\kappa }}_{{\bf{1u}}}}^{\bf{\dag }}{{\bf{\kappa }}_{{\bf{2u}}}}{a_1} + {{\bf{\kappa }}_{{\bf{2u}}}}^{\bf{\dag }}{{\bf{\kappa }}_{{\bf{2u}}}}{a_2}){{\bf{a}}_{\bf{t}}}^{\bf{\dag }}{{\bf{a}}_{\bf{t}}}}}{{{\gamma _2}{\gamma _m}}}.
\tag{A11}
\end{equation}
To optimize the efficiency, the targeted mode should follow $\frac{\partial }{{\partial {{\bf{a}}_{\bf{t}}}}}\eta  = 0$, which is given by
\begin{equation}
\frac{\partial }{{\partial {{\bf{a}}_{\bf{t}}}}}\eta  = \frac{{\partial \eta }}{{\partial {a_m}}}\frac{{\partial {a_m}}}{{a{{\bf{a}}_{\bf{t}}}}} + \frac{{\partial \eta }}{{\partial {a_2}}}\frac{{\partial {a_2}}}{{a{{\bf{a}}_{\bf{t}}}}}.
\tag{A12}
\end{equation}
Taking Eqs. (A9), (A10), and (A11) into Eq. (A12), we can get that
\begin{equation}
\frac{\partial }{{\partial {{\bf{a}}_{\bf{t}}}}}\eta  =  - i\frac{{\partial \eta }}{{\partial {a_m}}}\frac{{{a_1}{{\bf{\kappa }}_{{\bf{1u}}}} + {a_2}{{\bf{\kappa }}_{{\bf{2u}}}}}}{{{\gamma _m}}} - 2\frac{{\partial \eta }}{{\partial {a_2}}}\frac{{{{\bf{\kappa }}_{{\bf{1u}}}}^{\bf{\dag }}{{\bf{\kappa }}_{{\bf{2u}}}}{a_1} + {{\bf{\kappa }}_{{\bf{2u}}}}^{\bf{\dag }}{{\bf{\kappa }}_{{\bf{2u}}}}{a_2}}}{{{\gamma _2}{\gamma _m}}}{{\bf{a}}_{\bf{t}}}.
\tag{A13}
\end{equation}
To make the Jacobian matrix zero, as the necessary but not sufficient condition, the ${{\bf{a}}_{\bf{t}}}$ must be proportional to ${{\bf{\kappa }}_{{\bf{1u}}}}$ and ${{\bf{\kappa }}_{{\bf{2u}}}}$, as
\begin{equation}
{{\bf{a}}_{\bf{t}}} \propto {c_1}{{\bf{\kappa }}_{{\bf{1u}}}} + {c_2}{{\bf{\kappa }}_{{\bf{2u}}}},
\tag{A14}
\end{equation}
where the coefficients ${c_1}$ and ${c_2}$ are coefficients to be determined as illustrated in the main text.

We use metamaterial to manipulate the coupling coefficients ${\kappa _{1m}}$ and ${\kappa _{2m}}$. As shown in the main text, we choose the targeted mode to be ${{\bf{a}}_{\bf{t}}} \propto \sin \theta {{\bf{a}}_{{\bf{t - Tx}}}} + \cos \theta {{\bf{a}}_{{\bf{t - Rx}}}}$. As the position of the Rx resonator will influence ${{\bf{\kappa }}_{{\bf{2u}}}}$, and ultimately, ${{\bf{a}}_{\bf{t}}}$. The metamaterial configuration, given by $\left( {{\omega _u} - {\omega _0}} \right)/{\omega _0}$ distribution of the unit cells, is dependent on both $d$ and $\theta$. We plot the configurations in a square grid of $\left( {d,\theta } \right)$ space in FIG. S1. 
\newline
\begin{figure}[htbp]
\includegraphics[width=0.7\textwidth]{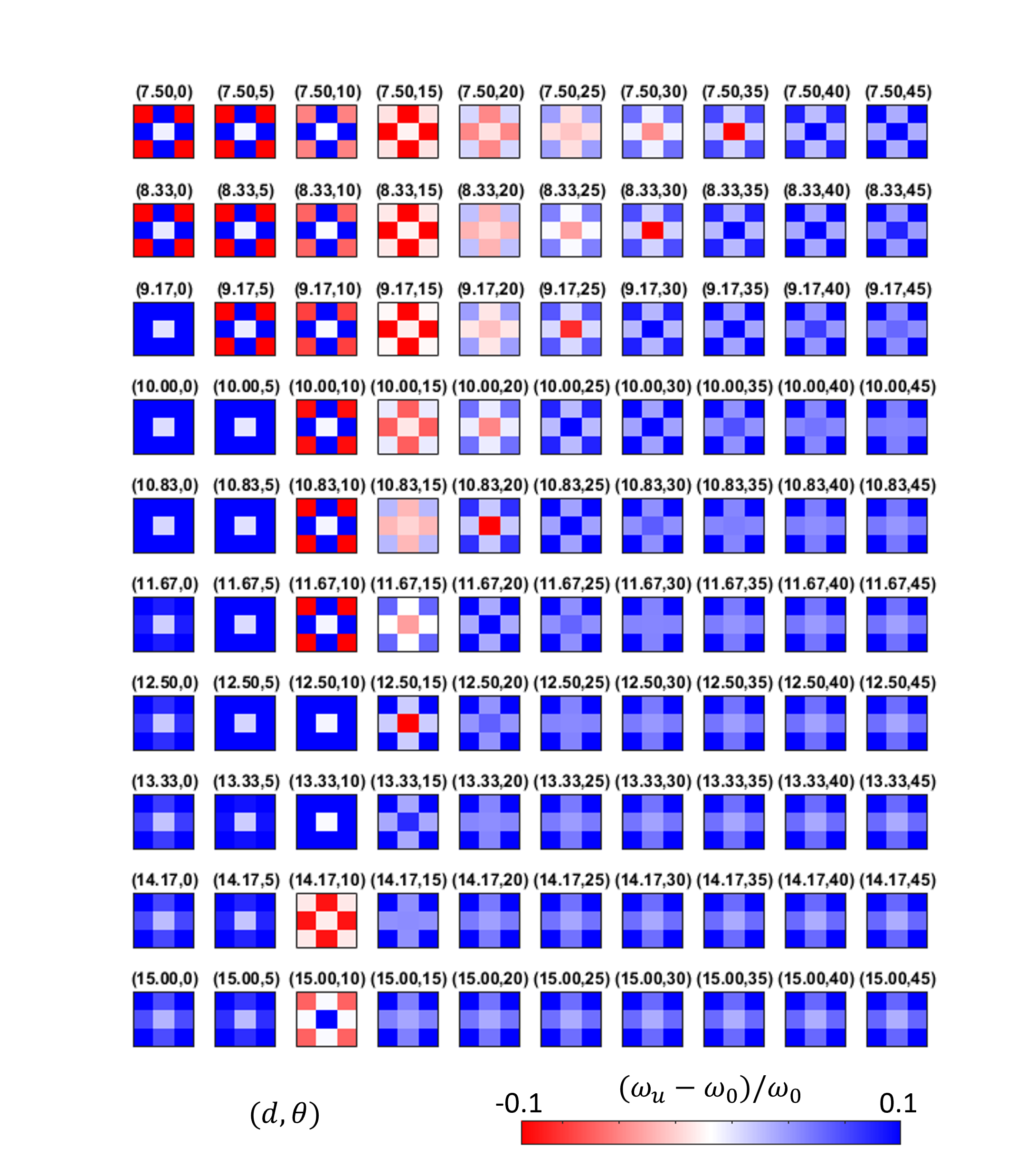}
\caption*{FIG. S1. Metamaterial’s configuration at different $d$ and $\theta$.}
\end{figure}

As the centers of the Tx resonator, the metamaterial, and the Rx resonator are aligned, the configurations are symmetric among the center of the metamaterial. As shown in FIG. S2, the configuration can be uniquely given by the differential frequency of the center unit cell (unit cell 5 in FIG. S2) with its surrounding unit cells. The surrounding unit cells can be separated into two groups: unit cells 2, 4, 6, 8 (FIG. S2(a)) and 1, 3, 7, 9 (FIG. S2(b)). The metamaterial configuration is uniquely reflected by   of the surrounding unit cells in these two groups. We plot $\left( {{\omega _u} - {\omega _0}} \right)/{\omega _0}$ in $\left( {d,\theta } \right)$ space. By using this diagram, we can easily acquire the configuration at any given $\theta$ and $d$. And we can see that the exceptional point falls in the region with a monotonous dependency of $\left( {{\omega _u} - {\omega _0}} \right)/{\omega _0}$ with $\theta$ and $d$, making the adjustment of the configuration to locate the $\delta  = 0$ condition straightforward.

\begin{figure}[htbp]
\includegraphics[width=0.7\textwidth]{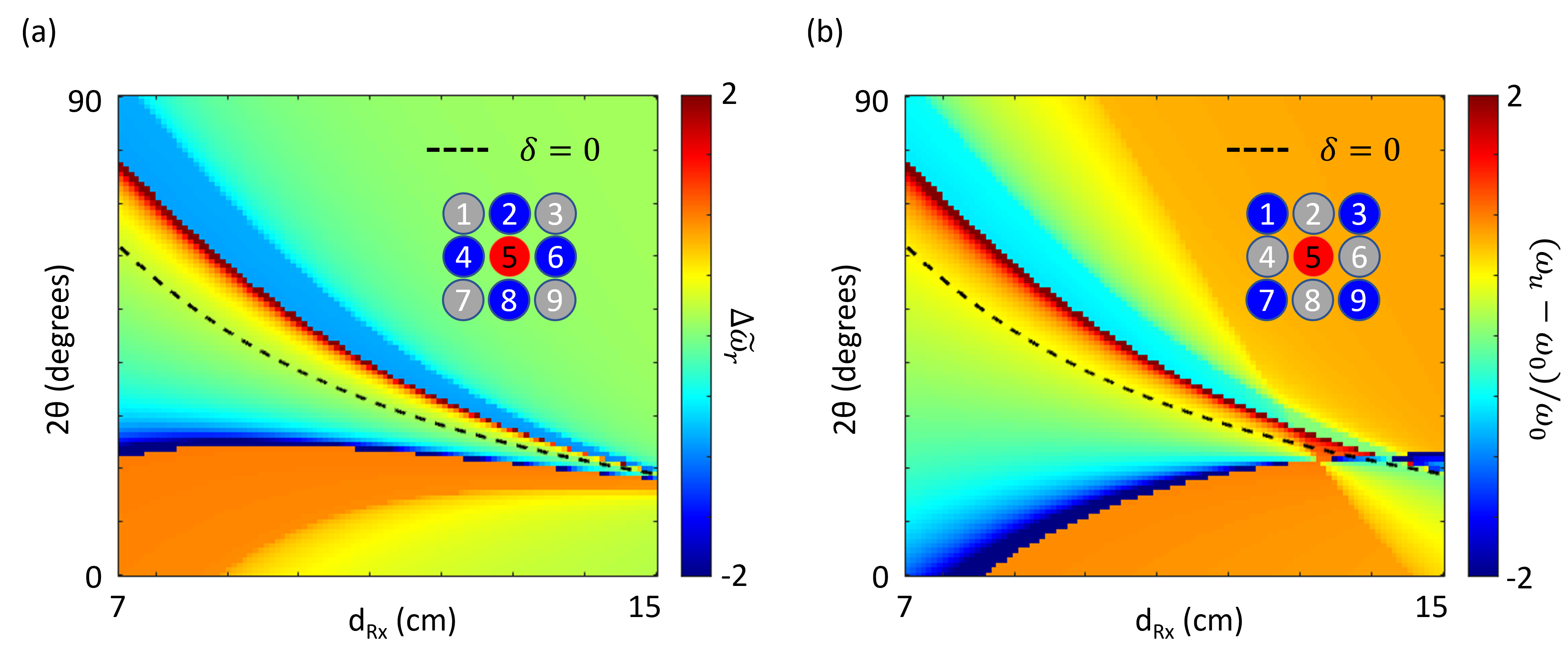}
\caption*{FIG. S2. Metamaterial configuration characterized by the differential unit cell resonance frequency, $\omega_u$. (a)$\left( {{\omega _u} - {\omega _0}} \right)/{\omega _0}$ of the center unit cell with its directly adjacent unit cells (marked in blue in the inset). (b) Differential $\left( {{\omega _u} - {\omega _0}} \right)/{\omega _0}$ of the center unit cell with the diagonal unit cells (marked in blue in the inset).}
\end{figure}

\subsection*{\Large Section 4. Transient analysis deriven by oscillation voltage input}
The temporal coupled-mode theory equation of the three-body system described by equation (3) of the main text is
\begin{equation}
\frac{d}{{dt}}{\bf{a}} = {{\bf{H}}_{3b}}{\bf{a}}.
\tag{A15}
\end{equation}
We calculate the coupling coefficient through mutual inductance. We assume a constant voltage, $v$, feeding to the Tx resonator, $g = \frac{v}{{{a_1}\sqrt {2{L_{Tx}}} }}$. Putting in the Hamiltonian, we get the standard time-dependent first-order ordinary differential equation (ODE) that can be solved numerically.
\begin{equation}
\frac{d}{{d\tilde t}}\left[ {\begin{array}{*{20}{c}}
{{a_1}}\\
{{a_m}}\\
{{a_2}}
\end{array}} \right] =  - i\left[ {\begin{array}{*{20}{c}}
1&{{k_1}}&{{k_0}}\\
{{k_1}}&{1 - i{\gamma _m}}&{{k_2}}\\
{{k_0}}&{{k_2}}&{1 - i\gamma }
\end{array}} \right]\left[ {\begin{array}{*{20}{c}}
{{a_1}}\\
{{a_m}}\\
{{a_2}}
\end{array}} \right] + \left[ {\begin{array}{*{20}{c}}
{v/\sqrt {2{L_{Tx}}} }\\
0\\
0
\end{array}} \right],
\tag{A16}
\end{equation}
where $\tilde t = {\raise0.7ex\hbox{${2\pi t}$} \!\mathord{\left/
 {\vphantom {{2\pi t} {{t_0}}}}\right.\kern-\nulldelimiterspace}
\!\lower0.7ex\hbox{${{t_0}}$}}$ represents normalized time in unit of the number of periods, ${t_0} = {\raise0.7ex\hbox{$1$} \!\mathord{\left/
 {\vphantom {1 {{f_0}}}}\right.\kern-\nulldelimiterspace}
\!\lower0.7ex\hbox{${{f_0}}$}}$. We set $\gamma  = 0.028$, ${k_0} = 0$, ${k_1} = 0.1$, ${k_2} = 0.1$, and ${\gamma _m} = 0$. As shown in FIG. S3, the scattering parameter, ${S_{21}} = \frac{{\left| {{a_2}} \right|}}{{\left| {{a_1}} \right|}}$, reaches saturation in around 4 periods. With an operating frequency of 65 MHz, the time of reaching saturation is around 61.6 ns. This means the stationary scattering parameter of the system is correlated with the PT symmetry of the state. Once forming the PT-symmetric state, the system will exhibit unity transmission property.

\begin{figure}[htbp]
\includegraphics[width=0.5\textwidth]{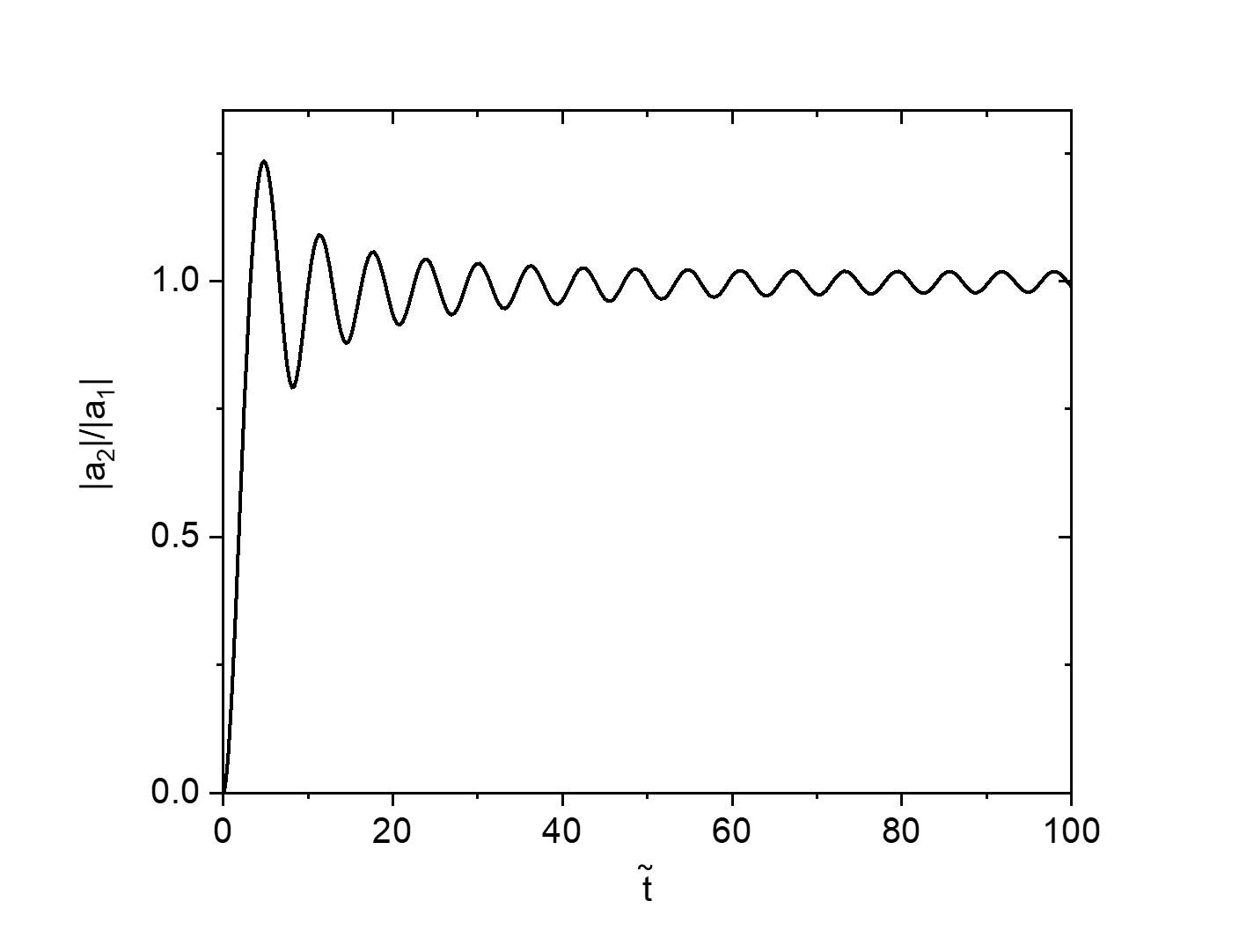}
\caption*{FIG. S3. Transient analysis under an oscillating voltage input.}
\end{figure}

\subsection*{\Large Section 5. Full wave simulation of the unit cell}
The resonance frequencies of the unit cells are simulated using COMSOL Multiphysics 6.0, AC/DC Magnetic Fields Module, and Frequency Domain solver. The cubic simulation domain has a diameter of 20 cm. We use constant magnetic field boundary conditions at all the 6 boundaries of the simulation domain. The feeding magnetic field is toward -z direct with an intensity of 1 A/m. The simulated magnetic field is shown in FIG. S4(a). As the resonator has open ends, the current density drops to zero at the two ends. The current density is highest at the middle point of the wire constructing the resonator and decrease with the distance to the middle point (FIG. S4(b)). As all the current flow to the same direction, the magnetic field distribution is similar to that of a coil.

\begin{figure}[htbp]
\includegraphics[width=0.7\textwidth]{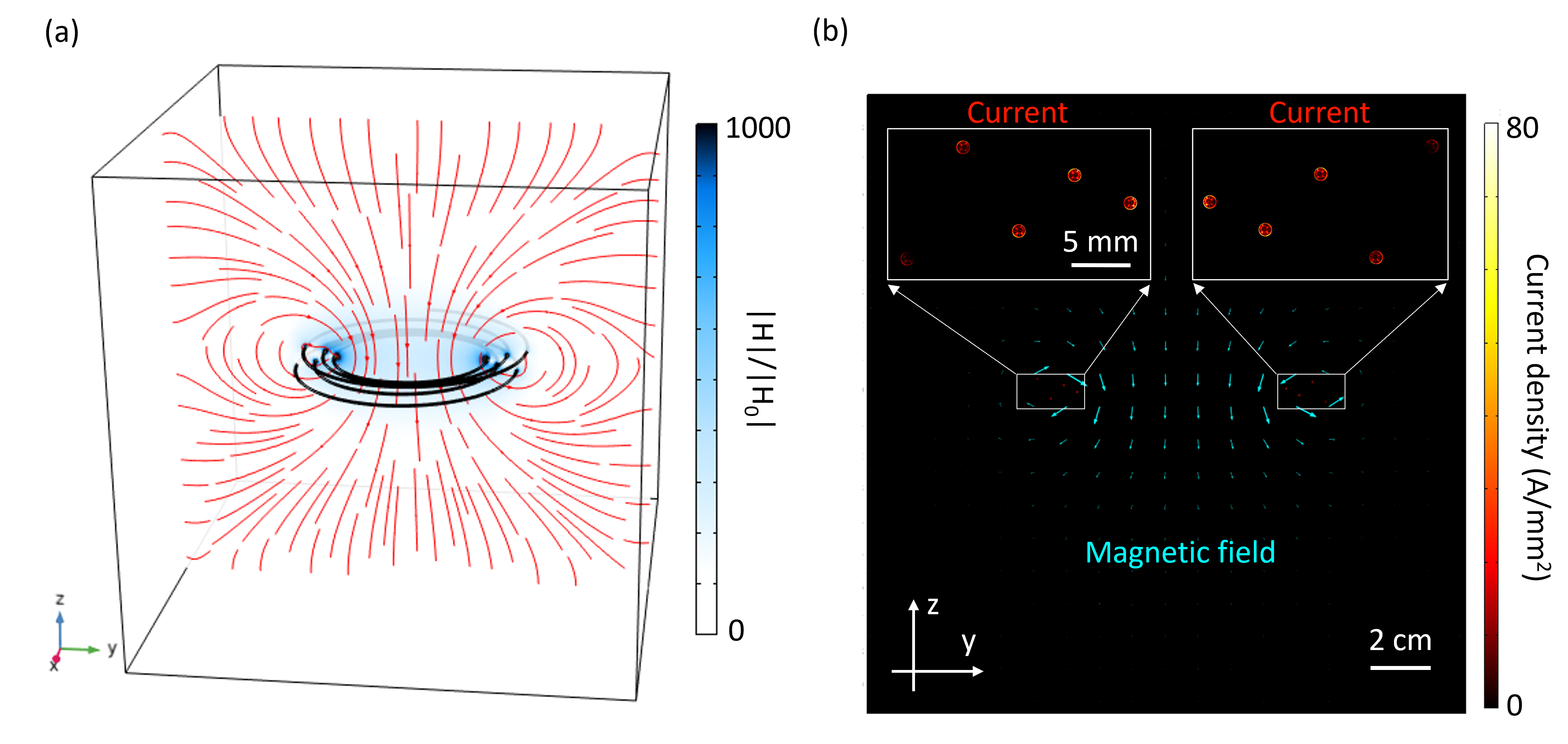}
\caption*{FIG. S4. Simulation of the unit cell. (a) Magnetic field intensity distribution. (b) Current density and magnetic field distribution. The unit cell has a height of 2 cm. The simulation frequency is 63.6 MHz.}
\end{figure}

\subsection*{\Large Section 6. Experimental setup}
The experimental setup is shown in FIG. S5. The external characterization system contains a waveform generator (Agilent 33250A Waveform Generator) and an oscilloscope (Agilent DSO-x 3034A). We use the waveform generator to drive the Tx coil using a linear sweep from 50 MHz to 80 MHz with a 10 volt peak-to-peak amplitude. The waveform generator sync is also connected to an input channel on the oscilloscope so that the time axis is related to the frequency sweep. We use another oscilloscope input channel to monitor the Rx voltage response versus frequency. 
\begin{figure}[htbp]
\includegraphics[width=0.7\textwidth]{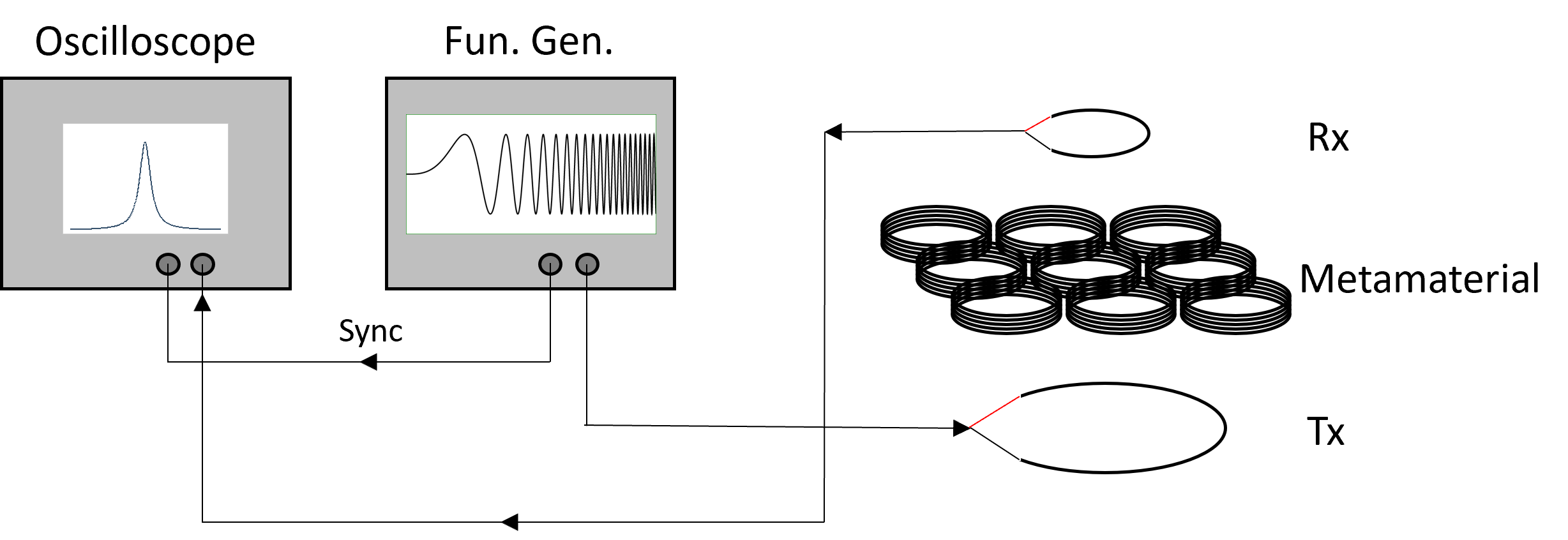}
\caption*{FIG. S5. Schematic of the measurement system.}
\end{figure}
\subsection*{\Large Section 7. System efficiency}
The efficiency of the system is given by
\begin{equation}
\eta  = \frac{{2{\gamma _L}{{\left| {{a_2}} \right|}^2}}}{{2{\gamma _{10}}{{\left| {{a_1}} \right|}^2} + 2{\gamma _{20}}{{\left| {{a_2}} \right|}^2} + 2{\gamma _L}{{\left| {{a_2}} \right|}^2}}}.
\tag{A17}
\end{equation}
According to Eq. (3) in the main text, we can get the relationship between the amplitudes $a_1$, $a_m$, and $a_2$ as
\begin{equation}
i\omega {a_1} = \left( {i{\omega _0} + {g_1}} \right){a_1} - i{\kappa _{1m}}{a_m}
\tag{A18}
\end{equation}

and

\begin{equation}
i\omega {a_2} = \left( {i{\omega _0} - {\gamma _2}} \right){a_2} - i{\kappa _{2m}}{a_m}.
\tag{A19}
\end{equation}
Combining Eqs. (A18)-(A19) and taking $\Delta \omega  = {\omega _i}i$, we can get the intensity ratio between ${a_1}$ and ${a_2}$ as
\begin{equation}
\frac{{{a_1}}}{{{a_2}}} = \frac{{{\kappa _{1m}}}}{{{\kappa _{2m}}}}\frac{{{\omega _i} - {\gamma _2}}}{{{\omega _i} + {g_1}}}.
\tag{A20}
\end{equation}
Take the amplitude ratio in Eq. (A20) to the efficiency in Eq. (A17), we get 
\begin{equation}
\eta  = \frac{{{\gamma _L}}}{{{\gamma _2} + {\gamma _{10}}\frac{{{\kappa _{1m}}^2}}{{{\kappa _{2m}}^2}}\frac{{{{\left( {{\omega _i} - {\gamma _2}} \right)}^2}}}{{{{\left( {{\omega _i} + {g_1}} \right)}^2}}}}},
\tag{A21}
\end{equation}
where ${\kappa _{1m}}^2 = {\kappa _{2m}}^2 + \delta $, under the strong coupling condition (${\gamma _2}^2 - s \le 0$), $\eta  = \frac{{{\gamma _L}}}{{{\gamma _2} + {\gamma _{10}}\left( {1 + \frac{\delta }{{{\kappa _2}^2}}} \right)\frac{{{{\left( {{\gamma _2} + \frac{{\left| \delta  \right|{\gamma _2}}}{{s - {\gamma _2}^2}}} \right)}^2}}}{{{{\left( {{g_1} - \frac{{\left| \delta  \right|{\gamma _2}}}{{s - {\gamma _2}^2}}} \right)}^2}}}}}$. Therefore, we can see that the efficiency is optimized as $\frac{{{\gamma _L}}}{{{\gamma _2} + {\gamma _{10}}\frac{{{\gamma _2}}}{{{g_1}}}}}$ when the coupling coefficients are balanced, $\delta  = 0$. Otherwise, the efficiency decreases as $\left| \delta  \right|$ increases.